\documentclass{iau}

\usepackage{amsmath}
\usepackage{natbib}
\usepackage{graphicx}
\usepackage{multirow}

\begin{document}

\lefttitle{Hugues Sana}
\righttitle{Interferometry of Massive Stars: Multiplicity, Magnetism, and Stellar Winds}

\jnlPage{1}{7}
\jnlDoiYr{2021}
\doival{10.1017/xxxxx}

\aopheadtitle{Proceedings IAU Symposium}
\editors{A. Wofford,  N. St-Louis, M. Garcia \&  S. Sim\'on-D\'iaz, eds.}

\title{Interferometry of Massive Stars: \\ Multiplicity, Magnetism, and Stellar Winds}

\author{H. Sana$^1$, E. Bordier$^2$, K.Deshmukh$^1$, A.J. Frost$^3$, A. Keskar$^1$, C. Lanthermann$^4$, R. R. Lefever$^{5}$, L. Mahy$^{6}$, A.A.C.  Sander$^5$, T. Shenar$^7$, F. Tramper$^8$}
\affiliation{1. Institute of Astronomy, KU Leuven, Celestijnlaan 200D, 3001 Leuven, Belgium. \\Email: \email{hugues.sana@kuleuven.be}}
\affiliation{2. I. Physikalisches Institut, Universitat zu Köln, Zulpicher Str. 77, Cologne, 50937, Germany}
\affiliation{3. European Southern Observatory, Alonso de Córdova 3107, Casilla 19, Santiago, Chile}
\affiliation{4. The CHARA Array of Georgia State University,         Mount Wilson Observatory, Mount Wilson, CA 91023, USA}
\affiliation{5. Zentrum für Astronomie der Universität Heidelberg, Astronomisches Rechen-Institut, Mönchhofstr. 12-14, 69120 Heidelberg, Germany}
\affiliation{6. Royal Observatory of Belgium, Avenue Circulaire/Ringlaan 3, B-1180 Brussels, Belgium}
\affiliation{7. The School of Physics and Astronomy, Tel Aviv University, Tel Aviv
6997801, Israel}
\affiliation{8. Centro de Astrobiología (CAB), CSIC-INTA, Carretera de Ajalvir km 4, E-28850 Torrejón de Ardoz, Madrid, Spain}

\begin{abstract}
After decades of efforts, optical long-baseline interferometry has become a mainstream observational technique in terms of operation robustness and user friendliness. Interferometry has opened a new observational window, enabling (sub)au-scale resolution of massive stars and direct measurements of orbital parameters, wind structures, and magnetic phenomena. This paper reviews recent advances in interferometric studies of massive stars, focusing on multiplicity, magnetism, and stellar winds.
\end{abstract}


\maketitle

\section{Introduction}
Massive stars (\(M > 8\,M_\odot\)) dominate the feedback processes in galaxies, contributing to chemical enrichment, ionizing radiation, and mechanical energy injection. Their short lifetimes and violent deaths as supernovae or gamma-ray bursts make them critical for understanding cosmic evolution. However, their evolution is far from simple: binarity, rotation, and magnetic fields introduce complex interactions that alter lifetimes, nucleosynthesis, and end states. Historically, multiplicity studies relied on photometry and spectroscopy, which are sensitive to short-period systems but blind to wider binaries. Adaptive optics and speckle interferometry have extended the reach to  separations of several 100 au. Yet in the early 2000s, a gap remained for intermediate separations of a few astronomical units \citep{Mason1998}.  Long-baseline, near-infrared and optical interferometry now bridges this gap, providing milli-arcsecond resolution and revealing itself as a discovery machine for new binaries and multiple systems \citep{Sana2014,Lanthermann2023,Frost2024}. 

\begin{figure}
\centering
\includegraphics[width=14cm]{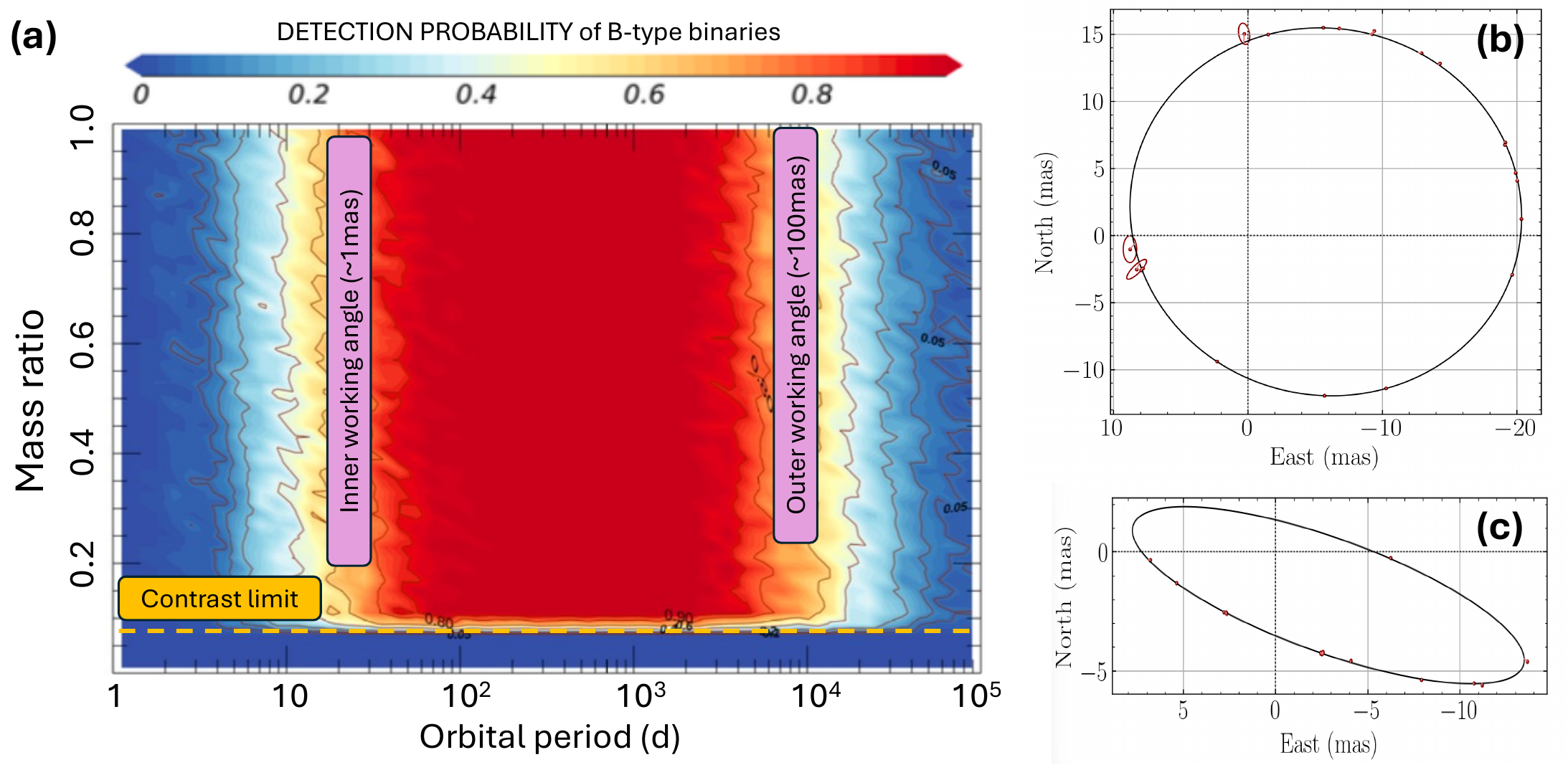}
\caption{(a) Binary detection probability of an interferometric campaign towards galactic early B-type stars (average distance 0.5~kpc); adapted from \citet{Frost2025}. (b) and (c) Relative astrometric orbits of $\delta$~Cir and MY~Ser (adapted from Sana+, in prep.)} \label{f:Pdetect}
\end{figure}

The probability of most techniques to successfully detect a binary system depends on temporal or spatial coverage, sensitivity to various flux or mass ratios, and/or the orbital phase and orbital configuration of the binary system itself. In the broad parameter space covered by binary systems \citep{Sana2026},  interferometry complements spectroscopy and photometry:  spectroscopy excels at short periods ($\lesssim 1~$year), interferometry probes intermediate regimes ($10^2$~days{-}$10^3$\ years) and adoptive-optics and other high-angular imaging techniques cover the widest separations \citep{SanaEvans2011}. This synergy is essential for mapping the full multiplicity parameter space. The large degree of homogeneity of interferometric surveys within the sensitivity range of the instrument is particularly noteworthy in the context of the present focus on interferometry (Fig.~\ref{f:Pdetect}a).

\section{Multiplicity of Massive Stars}
Ten years ago, the SMaSH+ survey increased the number of resolved companions from 64 to 260, confirming that the average O star hosts \( N_{\rm comp} = 2.2 \pm 0.3 \) companions \citep{Sana2014}. Recent efforts have transformed the observational measurements, in physical units, confirming an almost uniform mass-function in the interferometric range, but a lack of (similar)mass companions beyond about 500 to 1000~au \citep{Tramper2026}. A recent study of early-B stars revealed a fraction of multiple systems of $96 \pm 4\%$ and a companion frequency of $2.4 \pm 0.3$, questioning whether the accepted drop of multiplicity with lower stellar mass already starts occurring in the early B-star regime \citep{Offner2023}.

\section{Compact Massive Triples}

Within the 106 stars observed in the main sample of SMASH+, 51 have a resolved, interferometric companion. 21 of these also have an unresolved, closer companion, making them  hierarchical triple systems. Triples architectures influence mass transfer and merger rates, shaping populations of compact objects and gravitational-wave sources. \citet{Bordier2026} has showed that the triples are very hierarchical, with an outer to inner semi-major axis ratio in the range of 20-80, i.e.\ well beyond classical stability criteria  (Fig.~\ref{f:triples}a). This pile up at short period may results from dynamical processing through Kozai-Lidov cycle. Of interest, no additional correlation between separation, masses and mass ratio were found. While this may result from the small sample size, the sample is large enough for any strong correlation to be detectable.

Long-term follow up is  producing  the first orbits of the outer companions  (Figs.~\ref{f:Pdetect}b and c),.  Direct measurements of the mutual inclination $i_\mathrm{mut}$ between the orbital planes of the inner and outer orbit is given by
\begin{equation}
\cos i_\mathrm{mut}= \cos i_\mathrm{in} \cos i_\mathrm{out} + \sin i_\mathrm{in} \sin i_\mathrm{out} \cos(\Omega_\mathrm{in}-\Omega_\mathrm{out})
\end{equation}
Unfortunately, the longitude of the ascending node $\Omega_\mathrm{in}$ remains inaccessible without resolving the inner orbit. Given the high compactness of the inner orbits ($P_\mathrm{in} <30$~days), this will require the next generation of interferometric facilities with km-baselines. Despite the incomplete set of constraints, strong constraints on $|i_\mathrm{out}-i_\mathrm{in}|<1^o$ for $\delta$~Cir leaves this systems as a candidate coplanar system, while coplanarity is excluded for MY~Ser given $|i_\mathrm{out}-i_\mathrm{in}|= 82.4\pm0.3^o$ . Table~\ref{t:orbits} provides the interferometric orbital solution of both systems, and illustrates the extreme precision on total masses that relative astrometry provides.

\section{Magnetism and Binary Interaction}
\begin{figure}[t!]
\includegraphics[width=14cm]{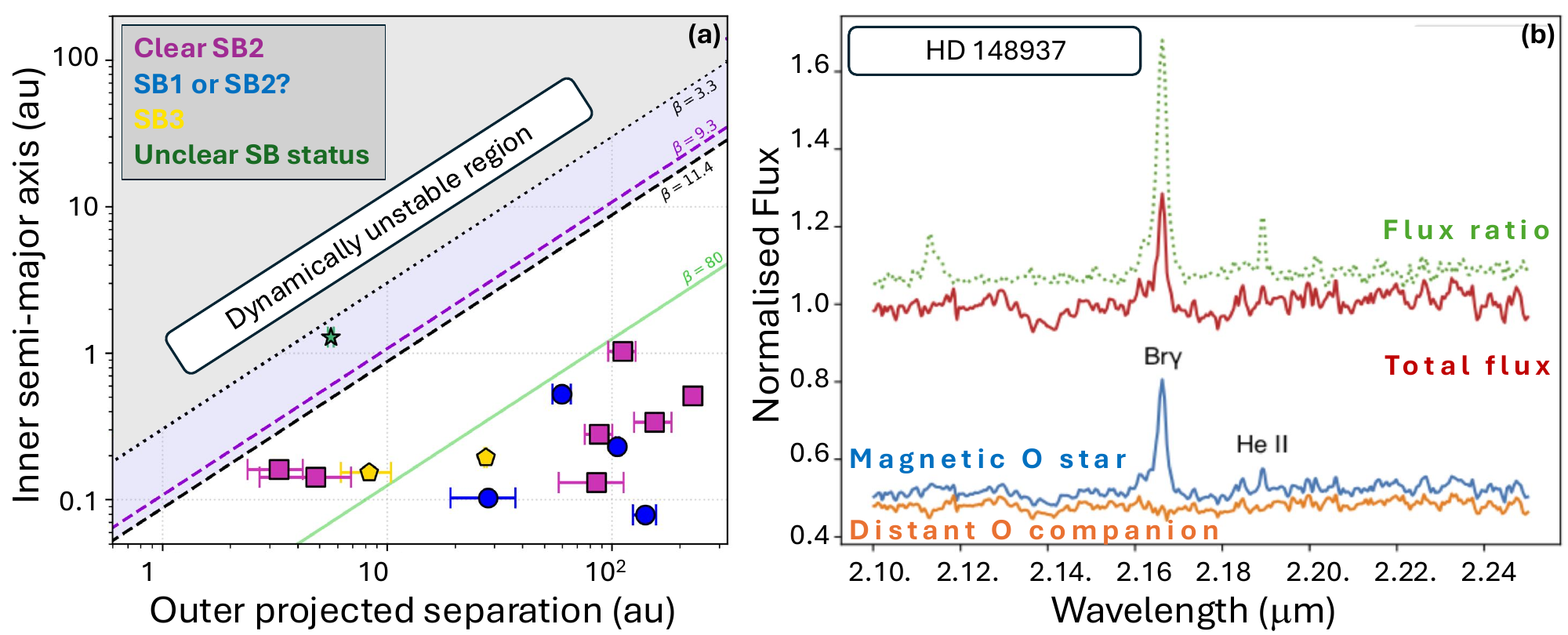}
\caption{(a) Inner {\it vs.} outer (projected) separation diagram of triple systems in the SMaSH+ sample. (b) Resolved $K$-band VLTI/GRAVITY spectra of the magnetic binary system HD~148937. Panels (a) and (b) are adapted from \citet{Bordier2026} and \citet{Frost2024}.  }
\label{f:triples}
\end{figure}

Magnetic fields add another layer of complexity. About 7\% of massive stars exhibit strong (\(>1\) kG) fossil magnetic fields \citep{Fossati2015, Wade2016}. These fields are stable, dipolar, and not generated by active dynamos, suggesting a primordial or merger origin. Simulations indicate that merger products rotate slowly and may evolve into magnetars \citep{Schneider2019}. Observationally, magnetic stars are rarely found in close binaries, supporting the hypothesis that mergers create magnetic fields. Case studies illustrate this link:
\begin{enumerate}
\item[$\bullet$] {\bf HD~148937} is a magnetic O+O binary with a long-period  ($P \sim 26$ yr) and a bipolar nebula~\citep{Frost2024}. Interferometry with VLTI/PIONIER and GRAVITY played a crucial role in identifying the magnetic star in the pair (Fig.~\ref{f:triples}b), and in obtaining strong constraints on the total mass ($e\approx0.78$, $M_\mathrm{ tot}=56.5 \pm 0.8$\,M$_\odot$).  Combined with the semi-amplitude of the non-magnetic massive star, total mass constraints can be converted in individual dynamical mass measurements. Together with spectral analysis, the physical characterization of the system components reveal a significant age discrepancy, where the magnetic merger product has been rejuvenated compared to its outer companion.  

\item[$\bullet$] {\bf HD~45166} is recently identified as a helium-rich star with a strong magnetic field, suggested to  be a magnetar progenitor \citep{Shenar2023}.  \citet{Deshmukh2025} used VLTI to resolved the system, providing assumption free mass-measurements, henceforth confirming the results of \citet{Shenar2023}.
\end{enumerate}

\begin{table}[t!]
    \centering
\caption{Astrometric orbital solution of the outer companion of the $\delta$~Cir and MY~Ser triple systems.} \label{t:orbits}
    \begin{tabular}{l c c c}
\hline
\hline
Parameter          &\multicolumn{2}{c}{$\delta$~Cir }       & MY~Ser   \\ 
\hline
$P_\mathrm{out}$ (d)          &\multicolumn{2}{c}{1603.20(63)}       & 6294.7(6.9)   \\ 
$e_\mathrm{out}$              &\multicolumn{2}{c}{0.510(3)}          &   0.423(1)    \\ 
$T_\mathrm{out}$ (MJD)        &\multicolumn{2}{c}{60\,499.1(2.0)}    &60\,894.72(99) \\ 
$i_\mathrm{out}$ ($^o$)       &\multicolumn{2}{c}{77.95(0.11)}       & 156.29(26)    \\
$\omega_\mathrm{out}$ ($^o$)  &\multicolumn{2}{c}{300.82(41)}        &109.84(49)     \\ 
$\Omega_\mathrm{out}$ ($^o$)  &\multicolumn{2}{c}{76.680(98)}        &218.28(52)     \\ 
$M_\mathrm{tot}$ (M$_\odot$)  &\multicolumn{2}{c}{33.20(48)}         & 69.90(22)     \\
$d$ (kpc; fixed)              &\multicolumn{2}{c}{0.704}               & 1.77          \\
\hline
\end{tabular}
\end{table}

These two examples not only highlight the likely role of binarity in shaping magnetic phenomena and exotic remnants, but also the crucial constraints that have been obtained from interferometry. In this context,  interferometry also played a key role in confirming the  Be-star + bloated stripped star nature of the HR~6819 system  proposed by \citet{Bodensteiner2020}.  Using VLTI, \citet{Frost2022} resolved the system for the first time while \citet{Klement2025} obtained the first 3D orbital solutions. The precision of the obtained masses ($M_\mathrm {Be}=4.03 \pm0.34$\,M$_\odot$, $M_\mathrm{stripped}=0.27 \pm0.06 $\,M$_\odot$) and corresponding mass-ratio ($M_\mathrm{Be}/M_\mathrm{stripped}=15.6\pm3.0$) are  challenging our ability to model this post-mass transfer system \citep{Picco2025}.

\section{Wolf-Rayet stars}
Wolf-Rayet stars are massive stars with strong, optically-thick winds which give rise to emission-line-dominated spectra and strongly influence the feedback and final fate  of massive stars. (A subset of) Wolf-Rayet binary systems with a distant O star companion are expected to be the outcome of Case B Roche-Lobe overflow \citep{Langer2020}, and an immediate evolutionary stage before the formation of single-degenerate binaries with a BH companion. An interferometric survey of 39 WRs accessible with the VLTI/GRAVITY however failed to reveal a significant number of companions in that range \citep{Deshmukh2024}. 

\begin{figure}[t!]
\includegraphics[width=14cm]{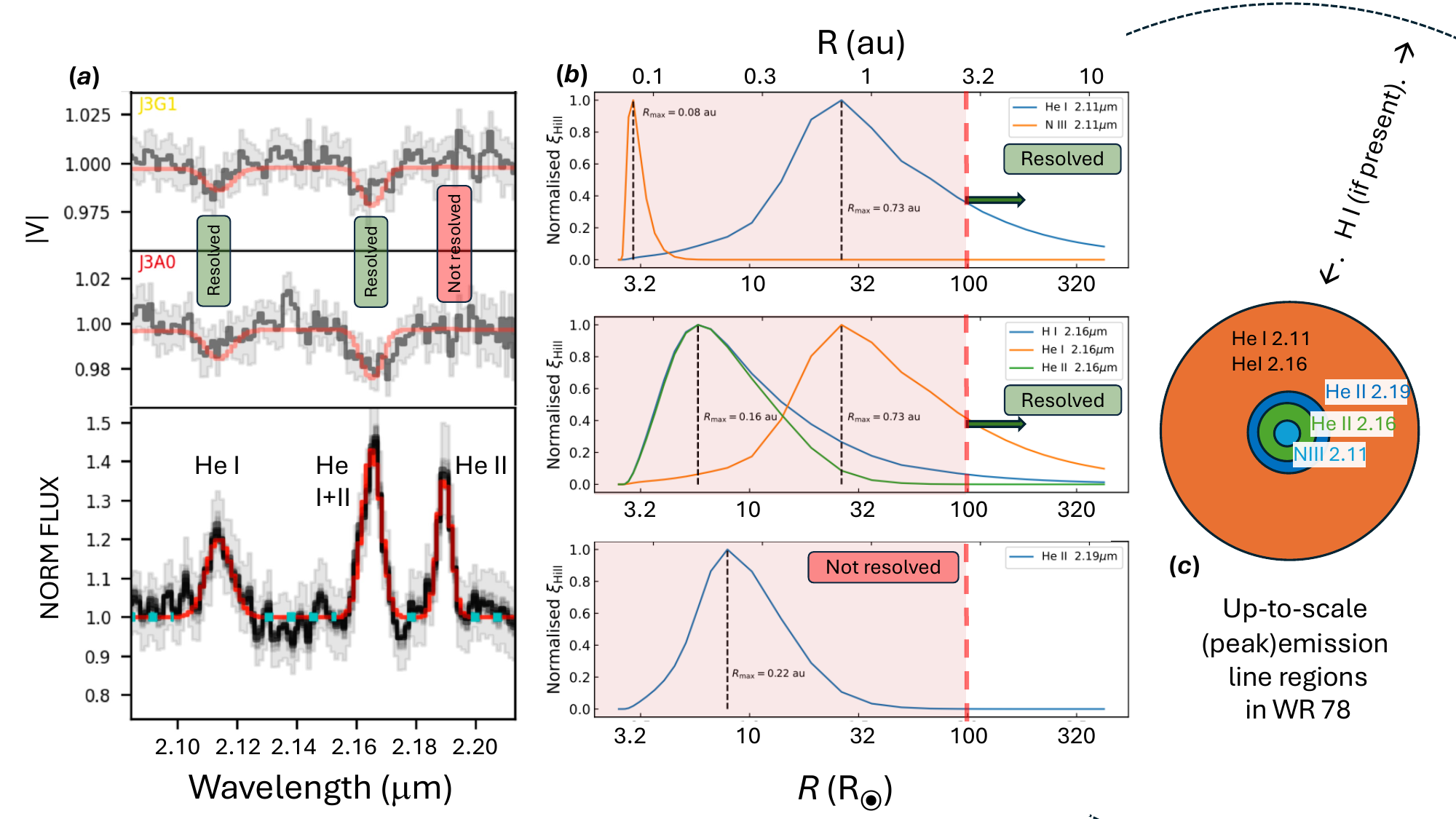}
\caption{Resolving wind line emission regions in WR~78. (a) Interferometric observables of  VLTI/Gravity observations. Bottom: normalized flux, revealing the presence of strong He~I and He~II lines. Top: closure phase spectra along two baselines, revealing a drop at the location of the He~I~2.11-2.16$\mu$m lines, but not at that of the He~II~2.19$\mu$m line, suggesting that the latter is unresolved while the former are (at least partially) resolved. (b) Predicted line formation strengths of various lines (see legend) in the wind of the WR78, computed with the PoWR atmospheric model \citep{powr2015,powr2002,powr2003}. The approximate inner working angle limit of the Gravity observation is indicated by the vertical dashed line. (c) Up-to-scale sketch of the relative size of the peak emission radii of the different lines in panel (b). Panels (a) and (b) are adapted from \citet{Deshmukh2024}.  }
\label{f:wr78}
\end{figure}

 As a serendipitous results of VLTI/GRAVITY observation, \citet{Deshmukh2024} showed that current interferometric capabilities start to resolve wind emission regions of several Wolf-Rayet stars, constraining the outer edge of line formation zones of He~I (resolved), and  He~II (unresolved) lines (see Fig.~\ref{f:wr78}).

\section{Conclusions}
The synergy of interferometry with other techniques enables full coverage of the parameter space relevant for massive binaries. These observations reveal that multiplicity is nearly universal, providing exquisite total mass constraints given enough orbital coverage. Similarly, interferometry has been providing unique constraints to identify the nature of binary products, such a merger or post-mass transfer objects. Interferometry was also crucial in confirming the  link between  (some) magnetic massive stars and a merger history. Looking to the future, kilometer-baseline arrays would bring another set of constraints, including critically resolving the inner binaries of triple systems and probing the winds of massive stars to an unprecedented level of accuracy.

\section*{Acknowledgments}
The research leading to these results has received funding from the European Research Council (ERC) (grant agreement numbers 772225: MULTIPLES) and  from the Flemish Government under the long-term structural Methusalem funding program, project SOUL: Stellar evolution in full glory, grant METH/24/012 at KU Leuven.

\bibliographystyle{plainnat}

\end{document}